\documentclass[aps,prb,superscriptaddress,papersize=a4paper,twocolumn,10pt,showpacs]{revtex4}
\usepackage{amsmath,amsfonts,amssymb,amsthm}
\usepackage{mathtext,mathtools,mathrsfs, bm, comment}
\usepackage{esint,bm,esvect,dsfont}
\usepackage{graphicx,epsfig,psfrag,xcolor}
\usepackage{array,tabularx,tabulary,booktabs,multirow}
\graphicspath{{figures/}}
\usepackage{hyperref}
\usepackage{verbatim}
\usepackage{float}

\def\ve{\varepsilon}

\def\vf{\varphi}
\def\de{\partial}

\def\bk{\mathbf{k}}
\def\bn{\mathbf{n}}
\def\bq{\mathbf{q}}
\def\bA{\mathbf{A}}

\def\de{\partial}

\def\ve{\varepsilon}

\def\vf{\varphi}

\def\br{\mathbf{r}}

\def\bp{\mathbf{p}}
\def\bn{\mathbf{n}}
\def\bq{\mathbf{q}}
\def\bk{\mathbf{k}}

\begin{document}

\title{Effects of anisotropy on the high field  magnetoresistance of Weyl semimetals}	
        \author{A.~S.~Dotdaev}
\affiliation{National University of Science and Technology MISIS, Moscow, 119049 Russia}
	\author{Ya.~I.~Rodionov}
		\affiliation{Institute for Theoretical and  Applied Electrodynamics, Russian Academy of Sciences, Moscow, 125412 Russia}
	\author{K.~I.~Kugel}
		\affiliation{Institute for Theoretical and  Applied Electrodynamics, Russian Academy of Sciences, Moscow, 125412 Russia}
\affiliation{National Research University Higher School of Economics, Moscow 101000, Russia}
    \author{B.~A.~Aronzon}
	\affiliation{P.~N.~Lebedev Physical Institute, Russian Academy of Sciences, Moscow, 119991 Russia}
	
\date{\today}

\begin{abstract}
We study the effects of anisotropy on the magnetoresistance of Weyl semimetals (WSMs) in the ultraquantum regime. We utilize the fact that many Weyl semimetals are approximately axially anisotropic. We find that anisotropy manifests itself in the strong dependence of the magnetoresistance on the polar and azimuthal angles determining the orientation of the anisotropy axis with respect to the applied magnetic field and electric current. We also predict that the ratio of magnetoresistances in the geometries, where the magnetic field and anisotropy axes are aligned and where they are orthogonal, scales as $(v_\bot/v_\parallel)^2$ where $v_\bot$ and $v_\parallel$ are the corresponding Fermi velocities.
\end{abstract}

\maketitle
\section{Introduction}
Weyl~\cite{burkov2011,wan2011,XuScience2015,lv2015} and Dirac~\cite{Liu2014,Neupane2014,Borisenko2014,Jeon_2014} semimetals attract intense interest in recent years. Due to their \textit{relativistic} 3D Hamiltonian with the Fermi velocity playing the role of the speed of light, they exhibit intriguing transport properties, e.g., disorder driven phase transitions~\cite{fradkin1986critical, syzranov2015}, unusual topological phenomena, such as the existence of Fermi arcs, an open in momentum space surface states connecting Weyl fermions of opposite chiralities~\cite{XuScience2015a}, and finally, pronounced QED-type phenomena such as chiral anomaly~\cite{armitage2018weyl,burkov2018weyl, son2013chiral,parameswaran2014}. To some extent, the material is essentially a solid-state realization of QED physics.

\begin{figure}[t]
\centering
 \includegraphics[width=0.45\textwidth]{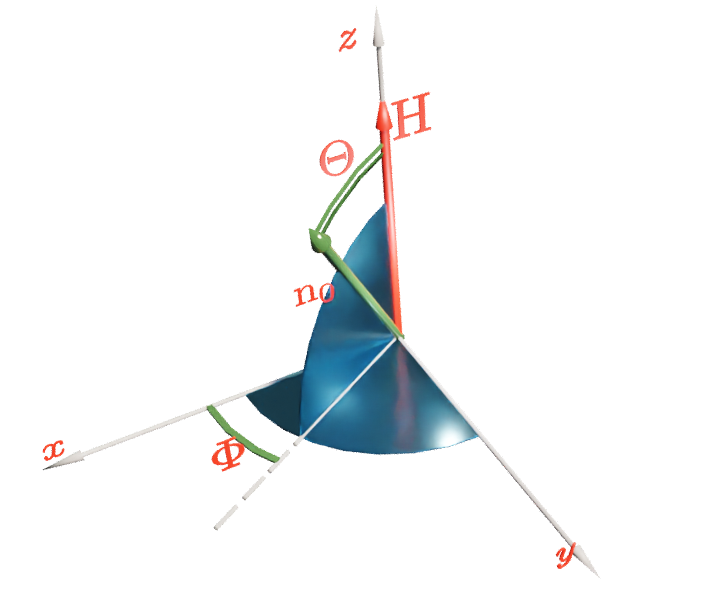}
 \caption{Geometry of the problem. The anisotropy axis ($\mathbf{n}_0$) is inclined by polar angle $\Theta$ and azimuth angle $\Phi$. The voltage is applied along the $x$ axis.}
\label{geom}
\end{figure}

Of particular interest are the transport properties of WSMs at the the magnetic field perpendicular to the transport voltage (transverse magnetoresistance). Recent experiments undertaken in the ultraquantum regime (at which temperature and chemical potential are much less than the energy gap between the zeroth and the first Landau levels (LLs)) reveal unsaturated magnetoresistance~\cite{liang2015ultrahigh, feng2015large, novak2015, zhao2015}, linear in the magnetic field $H$ ($\rho_{xx}\propto H$). As is, this behavior seems surprising since the usual relaxation-time arguments predict the saturation of the magnetoresistance at high magnetic fields. However, the transverse magnetoresistance of the compound with massless Dirac spectrum in the ultraquantum regime was theoretically studied by A. Abrikosov~\cite{Abrikosov1998} back in 1998. He assumed the principal source of the disorder in the compound to be Coulomb impurities. He found that magnetoresistance obeys linear law as a function of the magnetic field $H$.  In his work A. Abrikosov addressed the simplest isotropic gapless semiconductor with the linear spectrum identical to the one of a Dirac semimetal.

Actual WSMs are highly anisotropic compounds. Fortunately, for the theoretical analysis, some of the most popular ones, such as Cd$_3$As$_2$~\cite{Liu2014} or Na$_3$Bi~\cite{Liu2014_2} are approximately axially anisotropic with similar Fermi velocity ratios: $\xi=v_\bot/v_\parallel\approx 4$ and untilted Weyl cones. Naturally, the anisotropy of the materials substantially complicates the theoretical study. Most theoretical works so far addressed the anisotropy in WSMs caused by a possible tilt of the Weyl node, the so called type II WSMs~\cite{soluyanov2015type,trescher2015quantum}. In the meantime, the anisotropy of WSM with untilted Weyl cones is expected to have a dramatic effect on the experimental study of transport phenomena.  Indeed, an active experimental interest has recently awaken to the implications of anisotropy of WSMs with untitled Weyl cones~\cite{zhao2015,dhakal2022observation,huang2023laser}.

The effect of anisotropy of the untilted Weyl cone on transport properties of WSMs with the Coulomb disorder has not been studied theoretically yet. We note the comprehensive work~\cite{LeePRB2017}, where the effects of chemical potential and temperature on magnetoresistance in an isotropic WSM with the Coulomb disorder were (although mostly numerically) addressed. Also of note is the exhaustive study of magnetoresistance of isotropic WSMs with $\delta$-correlated disorder~\cite{KlierPRB2017}. The effect of strong Coulomb disorder on the transverse magnetoresistance was addressed in Ref.~\onlinecite{rodionov2023quantum}. The effect of anisotropy on the transport of WSM with long-range disorder  without magnetic field was studied in Ref.~\onlinecite{rodionov2015effects}.

In this paper, we compute the magnetoconductivity and magnetoresistance of a WSM with axially anisotropic untilted Weyl cone in the ultraquantum regime. We obtain the magnetoresistance as a function of the magnetic field, and of the polar and azimuthal angles of the anisotropy axis (see Fig.~\ref{geom} for the actual geometry). We analyze the scaling of the conductivity tensor components with the anisotropy parameter $\xi = v_{\parallel}/{v_\bot}$.

The paper is organized as follows. In Sec. II, we introduce the anisotropic WSM Hamiltonian and discuss the transformation properties of conductivity necessary for the computation. Sec. III addresses the computation of the magnetoconductivity. In Sec. IV, we deal with the magnetoresistance and analyze its $\xi$ and angular dependence. We summarize the results of the paper in Sec. VI and discuss the regime, in which they are applicable.

\section{Formulation of the model}

\subsection{Hamiltonian}

We start with the standard anisotropic Hamiltonian for electrons in the Coulomb disorder potential
    \begin{align}
       H&=H_0+H_{\rm imp},\notag\\
       H_0&= \sum\limits_{i=\bot,\parallel}\int\psi^\dag(\br)\bm{\sigma}_i\left(v_i\Big[\mathbf{p}-\frac{e}{c}\mathbf{A}\Big]_i\right)\psi(\br) d\br, \label{ham1}\\
       H_{\rm imp} &= \int \psi^\dag(\br) u(\br)\psi(\br) d\br,  \label{ham_dis}
\end{align}
where $H_0$ is the anisotropic Hamiltonian of noninteracting Weyl fermions, $\psi(\mathbf{r})$ and $\psi^\dag(\mathbf{r})$ are the fermion annihilation and creation operators, $\bm{\sigma}_\parallel = (\bm{\sigma}\cdot\mathbf{n}_0)\bn_0,\ \bm{\sigma}_\bot =\bm{\sigma} - \mathbf{n}_0(\bm{\sigma}\cdot\mathbf{n}_0)$, are the Pauli matrices, $v_\parallel$ and  $v_\bot$  are the Fermi velocities, and $\mathbf{n}_0$ is the unit vector determining the direction of the anisotropy axis (see Fig.~\ref{geom}).  The term $H_{\rm imp}$ is responsible for the interaction between electrons and Coulomb impurities.
For the reference, we present the details of derivation of Hamiltonian~\eqref{ham1} in Appendix~\ref{App_ham}.

As is well known, the Nielsen--Ninomiya theorem~\cite{nielsen1981absence} states that the Weyl nodes should appear in pairs within the Brillouin zone. However, due to the smoothness of the disorder potential (see the details in the Discussion section), we discard the charge carrier scattering between the nodes. Therefore, to determine the full conductivity, one simply multiplies the result from a single Weyl node by the number of nodes in the Brillouin zone of the WSM.
Throughout the paper, we set $\hbar = 1$ and introduce the variable $\Omega$, related to the magnetic field  (the distance between the zeroth and the first LL) and the magnetic length $l_H$
\begin{gather}
\label{notation}
    \begin{split}
    \Omega^2 &= \frac{2eHv_\parallel}{c},\quad l_H^2  = \frac{c}{eH}.
    \end{split}
\end{gather}

\subsection{Disorder potential}
The screened disorder potential reads

\begin{gather}
\label{disorder_1}
    u(\mathbf{k}) = \frac{4\pi e^2}{\epsilon}\frac{1}{k^2-\frac{4\pi e^2}{\epsilon}\Pi(k^2)} ,
\end{gather}
where $\epsilon$ is the dielectric constant and $\Pi(k)$ is the Fermi gas polarization operator taken in the static limit (frequency is set to zero): $\Pi(k^2) \equiv \Pi(\omega, k^2)|_{\omega=0}$. In the situation of the ordinary Fermi liquid, the momentum transferred by the static disorder potential to a charge carrier is much smaller than the Fermi momentum $k\ll k_{\rm F}$. This entails the possibility to expand $\Pi(k^2)$ in terms of $k/k_{\rm F}\ll1$ in Eq.~\eqref{disorder_1}, and keep the first term only:
$\Pi(k) = \Pi(0)+k^2\de_{k^2}\Pi(0) +...$, where $\Pi(0) = -dn/d\mu$ is the thermodynamic density of states. This leads to the standard static screening of the Coulomb interaction.

In our problem, as we will see in the course of calculations, the situation is more subtle. The role of Fermi momentum is assumed by the inverse magnetic length $l_H^{-1}$.  We may write the expression for the exact polarization operator in the following suitable form
\begin{gather}
    \begin{split}
    \Pi(k^2) &= -\frac{dn}{d\mu}(1+c_1(kl_H)^2+c_2(kl_H)^4+...)\\
    &=-\frac{dn}{d\mu}[1+k^2l_H^2f(k^2l_H^2)],
    \end{split}
\end{gather}
where $f(0)\neq 0$ and $f(x)$ is some dimensionless function measuring a deviation of the polarization operator from its value at zero momentum. At low temperatures, only the zeroth Landau level is occupied and $dn/d\mu$ is easily calculated (see e.g. Ref.~\onlinecite{KlierPRB2015}) yielding $dn/d\mu = (2\pi^2 v l_H^2)^{-1}$. Using Eq.~\eqref{disorder_1}, we write the following expression for the screened disorder potential
\begin{gather}
\label{coulomb2a}
    \begin{split}
    u(\bk) &= \frac{4\pi e^2}{\epsilon}\frac{1}{k^2[1+\frac{2\alpha}{\pi} f(k^2l_H^2)]+\frac{2\alpha}{\pi l_H^2}},
    \end{split}
\end{gather}
where
\begin{gather}
    \alpha = \frac{e^2}{\epsilon\hbar v_{\parallel}},
\end{gather}
is the so called fine structure constant for WSM.

We will see that the main contribution to the conductivity related to the disorder potential comes from the  $k\lesssim l_H^{-1}$ momentum range, where $l_H$ is defined in Eq.~\eqref{notation} (see Appendix~\ref{main_log} for the details). As a result, in contrast to the Fermi liquid theory, the argument of function $f$ entering denominator of Eq.~\eqref{coulomb2a} is of the order of unity. Therefore, the whole expression is $f(k^2 l_H^2)\sim\mathcal{O}(1)$ in our problem.

However, as is known quite well, the typical WSM, like Cd$_3$As$_2$ has an additional small parameter $\alpha \ll 1$ which is for Cd$_3$As$_2$ is equal $\alpha\approx 0.05$~\cite{Liu2014,Aubin1977}. 
This drastically simplifies our analysis. Taking into account exact $\Pi(k^2)$ in the Coulomb disorder~\eqref{disorder_1} instead of $\Pi(0)$ is equivalent to keeping the term with function $f$ in expression \eqref{coulomb2a}. However, as one sees from \eqref{coulomb2a}, $f$ enters with small prefactor $\alpha$ in the renormalization of the Coulomb field.

Thus, keeping the term containing $f$ in Coulomb interaction \eqref{coulomb2a} yields small (of the order of $\alpha$) corrections to the observables. We, on our part, will keep only the terms of the order of $\mathcal{O}(\ln\alpha)$ and $\mathcal{O}(1)$. Therefore, we will substitute $\Pi(k^2)$ by $\Pi(0)$ in disorder potential~\eqref{disorder_1} and will use the standard Lindhard expression for the renormalized Coulomb potential
 \begin{gather}
 \label{coulomb2}
    u(\mathbf{k}) = \frac{4\pi e^2}{\epsilon}\frac{1}{k^2+\kappa^2},
\end{gather}
where $\kappa^2 = 2\alpha\pi^{-1} l_H^{-2}\ll l_H^{-2}$ is the inverse Debye screening length squared, from now on.

\subsection{Transformation of the conductivity tensor}
\label{cond_transform}

Before we proceed any further, it is quite suitable to introduce the rescaling, which makes the spectrum isotropic
\begin{gather}
\label{scaling}
 r_\parallel = r_{s,\parallel},\ \ r_\bot = \xi r_{s,\bot},\ \ \psi(\br) =
 \frac{1}{\xi}\psi_s(\br_s),\ \ v_\bot = \xi v_\parallel .
\end{gather}
Transformation~\eqref{scaling} makes the disorder-free part of the Hamiltonian isotropic
\begin{gather}
        \label{ham2}
    \begin{split}
       H_{s,0}&= -iv_\parallel\int\psi^{\dag}_s(\br_s)\bm\sigma\left(\nabla_{\br_s}-i\frac{e}{c}\mathbf{A}\right)\psi_s(\br_s) d\br_s,\\
       H^\prime_{\rm imp} &= \int \psi^{\dag}_s(\br_s) u(\br_{s,\parallel}+\xi\br_{s,\bot})\psi(\br_s) d\br_s.
    \end{split}
\end{gather}
The transformation is performed in three steps. First, we rotate the coordinate system so that the new $z^\prime$ axis becomes parallel to the anisotropy axis. We rotate it  by angle $\Phi$ about axis $z$ and then by angle $\Theta$ about the transformed $y$ axis (see Fig.~\ref{geom}).
\begin{gather}
\label{cond_rot1}
    \sigma  = R\sigma^\prime R^{-1},
\end{gather}
where matrix $R$ is presented in Appendix~\ref{App_ham}, Eq.~\eqref{rot}.
Second, we perform the rescaling. We denote the rescaled conductivity tensor in the rotated basis as $\sigma^\prime_{s}$. The correct transformation rule is not immediately obvious. The details are summarized in Appendix~\ref{App_cond_transform}. The transformation rule has the form
\begin{gather}
    \sigma^\prime = S_1\sigma^\prime_{s} S^{-1},
\end{gather}
where matrices $S$ and $S_1$ are defined by Eqs.~\eqref{matrix_S} and \eqref{current}

\begin{figure}[H]
\centering
\includegraphics[width=0.5\textwidth]{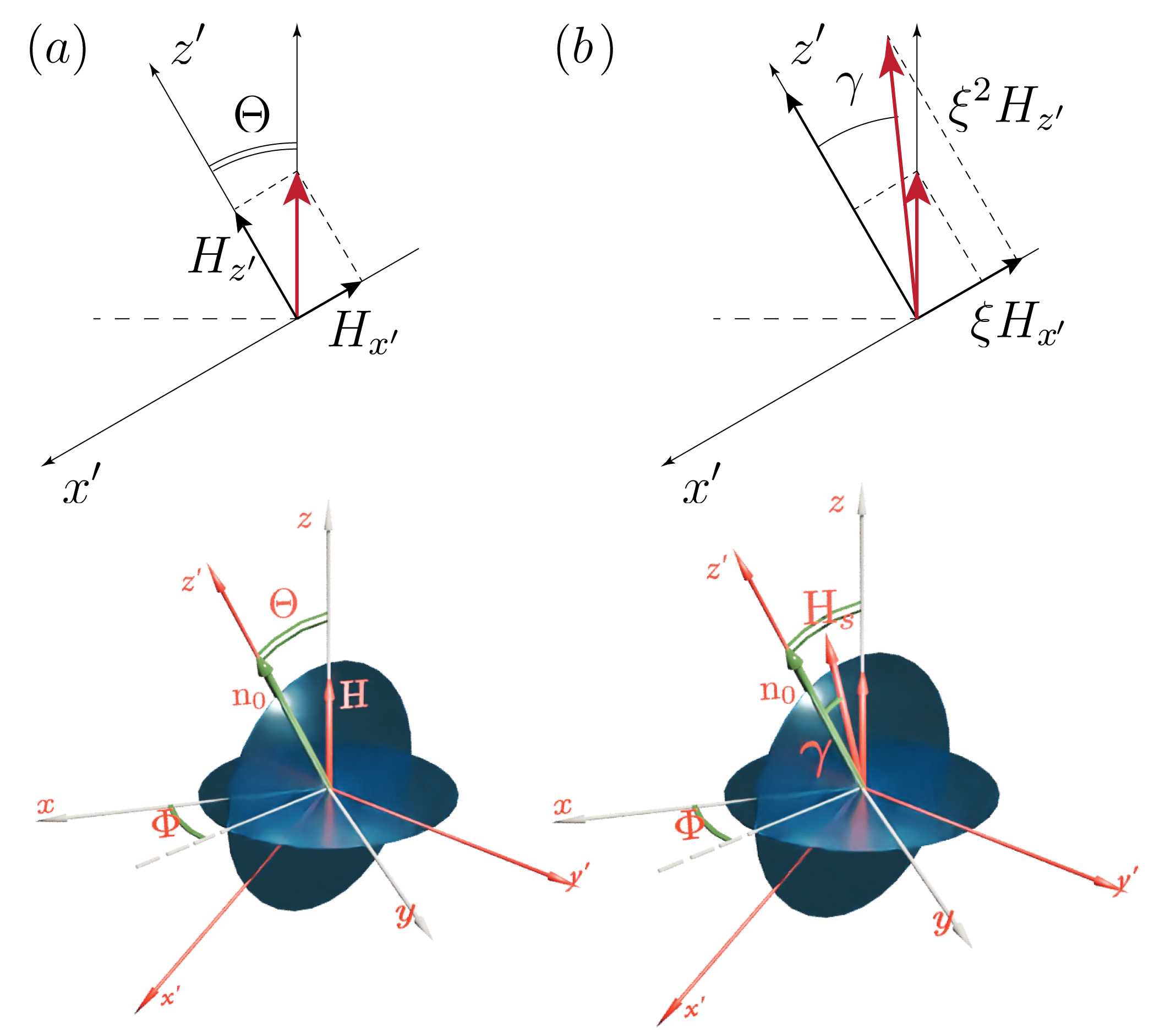}
 \caption{ (a) Rotated basis
  $x^\prime,y^\prime,z^\prime$. Axis $z^\prime$ is oriented along the anisotropy vector $\mathbf{n}_0$. (b) Position of the rescaled magnetic field vector $\mathbf{H}_s$ after the rescaling. We can see that it remains in the $z^\prime y^\prime$ plane, but is rotated by angle $\gamma$ about $y^\prime$ axis.}
\label{geom2}
\end{figure}

The scaling transformation also changes the components of the  magnetic field vector $\mathbf{H}$. The transformation law is derived in Appendix~\ref{App_cond_transform}, (see Eq.~\eqref{electric_field})
\begin{gather}
\label{magnetic_scaled}
    \mathbf{H}_s = H(-\xi  \sin\Theta,\ 0,\ \xi ^2 \cos \Theta).
\end{gather}
We see that the magnetic field changes according to the law
\begin{gather}
\label{magnet_str}
    H_s=\xi\eta H,\ \ \eta=\sqrt{\xi^2\cos^2\Theta+\sin^2\Theta}.
\end{gather}
It is important to note that Eq.~\eqref{magnetic_scaled} also entails the change of the inclination angle of the vector with respect to the scaled basis (though the direction of the vector, of course, stays unchanged. However, to accentuate the fact that the measured angle is changed, we draw vector $\mathbf{H}_s$ in a slightly different direction in Fig.~\ref{geom2} for illustrative purposes). As is seen in the figure, the rescaled magnetic field vector is inclined by an angle
\begin{gather}
    \gamma = -\arctan\left(\frac{1}{\xi}\tan\Theta\right)
\end{gather}
in the rotated $(x^\prime z^\prime)$ plane with respect to the $z^\prime$ axis. To make the calculation for the conductivity easier, we need to switch to the coordinate system, in which $z$ axis is aligned along the magnetic field vector. Therefore, we need to perform the reversed rotation by the $\gamma$ angle in the $(x^\prime z^\prime)$ plane (we denote the corresponding rotation matrix as $R_\gamma$). Let us denote the new conductivity tensor in the once more rotated basis as $\sigma^\prime_{s}$
\begin{gather}
\label{cond_fin_rot}
    \sigma_{s} = R_\gamma \sigma^\prime_{s} R_\gamma^{-1} ,
\end{gather}
where the $R_\gamma$ matrix is identical to matrix $R$ from Eq.~\eqref{rot} up to the change $\Theta\rightarrow\gamma,\ \Phi\rightarrow 0$.
As a result, the initial conductivity tensor and the rescaled and rotated one are related by the following transform
\begin{gather}
\label{cond_fin_rot1}
    \sigma = R S_1 R_\gamma \sigma^\prime_{s} R_\gamma^{-1} S^{-1} R^{-1}.
\end{gather}

This results in the following final expression for relating components of the conductivity tensor in the rotated rescaled basis and the initial one
\begin{gather}
\label{cond_fin_rot2}
    \begin{split}
    \sigma_{xx} &= \frac{1}{\xi^2}\big[\eta^2\sigma^\prime_{s,xx}\cos^2\Phi+
    \xi^2\sigma^\prime_{s,yy}\sin^2\Phi\big],
    \end{split}
\end{gather}
where $\eta$ is defined in Eq.~\eqref{magnet_str}.

We note here that the anisotropy axis is now inclined by the polar angle $\gamma$ in the $(x^\prime z^\prime)$ plane. The latter means that the axis's azimuthal angle is zero. The components of the conductivity tensor, in general, should depend on the Euler angles.

Next, we realize that the conductivity tensor components $\sigma^\prime_{s, xx}$ and $\sigma^\prime_{s, yy}$ ought to depend on the component of the vector determining the direction of the anisotropy axis (anisotropy vector). However, the only geometric difference between tensor components $\sigma^\prime_{s,xx}$ and $\sigma^\prime_{s,yy}$ is the orientation of the anisotropy vector with respect to $x^\prime y^\prime$ plane. Therefore, we have $\sigma^\prime_{s,yy}(\vf) = \sigma^\prime_{s,xx}(\pi/2-\vf)$, where $\vf$ is the azimuthal angle.

Also, we should pay attention to the behavior of conductivity \eqref{cond_fin_rot2} at $\Theta = 0$. In this case, the anisotropy axis coincides with the direction of the magnetic field. In such situation, the azimuthal angle $\Phi$ is, strictly speaking, undefined. Therefore, the conductivity tensor is supposed to be independent of $\Phi$ at $\Theta = 0$. As will be proven in the next section, at $\Theta = 0$ (anisotropy axis is aligned along the direction of the magnetic field), we obtain that $\sigma^\prime_{s,xx} = \sigma^\prime_{s,yy}$, and Eq.~\eqref{cond_fin_rot2} implies the relation $\sigma_{xx} = \sigma^\prime_{s,xx}$. The latter is quite natural since in this case, the system effectively becomes isotropic in the $(xy)$ plane.

\subsection{Debye screening}

The inverse Debye screening length is determined according to the standard equation
\begin{gather}
    \kappa^2 = \frac{4\pi e^2}{\epsilon}\frac{dn(H)}{d\mu},
\end{gather}
where $n$ is the particle density determined by the chemical potential $\mu$. The easiest way to compute the particle density at the applied magnetic field is to switch to the rescaled rotated basis. The rescaled density is related to the initial one via the transform: $n_s = \xi^{2}n$ (see discussion of Eq.~\eqref{current}). As a result, the Debye screening is determined as
\begin{gather}
    \kappa^2 = \frac{1}{\xi^2}\frac{dn_s(H_s)}{d\mu}\equiv
    \frac{1}{\xi^2}\frac{2\alpha}{\pi l_{H_s}^2},
\end{gather}
where $l_{H_s} = c/eH_s\equiv c/(eH\xi\eta)$ is the magnetic length in the rescaled coordinate system.

The rescaled disorder potential leads to the modified disorder correlation function
\begin{gather}
\label{dis_res}
    g(p) = \frac{16\pi^2 n_{\rm imp} \xi^2\alpha^2 v_\parallel^2}{(\xi^2p_\parallel^2+p_\bot^2 + \xi^2\kappa^2)^2}.
\end{gather}

Now, we are ready to compute the conductivity. To this end, we are going to employ the Kubo formalism. As usual, the conductivity contains two distinct contributions: the one, which comes from the separate averaging of Green's functions, and the vertex correction.

\section{Conductivity  $\sigma_{xx}$}
\subsection{Kubo expressions and Green's functions}

The expression for conductivity is given by the standard Kubo formula (see e.g. Ref. ~\onlinecite{KlierPRB2015})
\begin{widetext}
\begin{gather}
    \label{cond2}
    \begin{split}
   \sigma_{xx} &= 2e^2v_{\parallel}^2\int\frac{d\ve d\mathbf{p}dx^\prime}{(2\pi)^3}\frac{df(\ve)}{d\ve}{\rm Tr}
   \bigg[
        \langle{\rm Im}G_{11}^R(x,x^\prime;\ve,\bp)
        {\rm Im}G_{22}^R(x^\prime,x;\ve,\bp)
        \rangle +
    \langle{\rm Im}G_{22}^R(x,x^\prime;\ve,\bp)
        {\rm Im}G_{11}^R(x,x^\prime;\ve,\bp)
        \rangle\\
    &-\frac{1}{4}\langle\big[G_{12}^R(x,x^\prime;\ve,\bp)
    -G_{12}^A(x,x^\prime;\ve,\bp)\big]
        \big[G_{12}^R(x^\prime, x;\ve,\bp)
    -G_{12}^A(x^\prime, x;\ve,\bp)\big]
        \rangle\\
    &-\frac{1}{4}\langle\big[G_{21}^R(x,x^\prime;\ve,\bp)
    -G_{21}^A(x,x^\prime;\ve,\bp)\big]
        \big[G_{21}^R(x^\prime, x;\ve,\bp)
    -G_{21}^A(x^\prime, x;\ve,\bp)\big]
        \rangle
   \bigg] .
    \end{split}
\end{gather}
\end{widetext}
Here, angular brackets denote the disorder averaging, and $f(\ve)$ is the Fermi distribution function. The integration over momentum $\bp$ is performed in the $(p_y,p_z)$ plane. The last two lines in Eq.~\eqref{cond2} (usually absent in standard analysis) appear owing to the disorder vertex corrections and, as we will see below, do not vanish only in the anisotropic case.

The Green's functions entering~\eqref{cond2} are defined as follows
\begin{gather}
\label{green-zero}
    \begin{split}
    &G^R(x,x^\prime;\ve,\bp)= \sum\limits_{n=0}^\infty S_n(x_{p_y})G^R_n(\ve,\bp_n) S^\dag_n(x_{p_y}^\prime) ,\\
              &S_n(s)=\begin{pmatrix}
            \chi_n\big(s\big)\ &\ 0\\
            0\ &\ \chi_{n-1}\big(s\big)
           \end{pmatrix},\\
           &G^R_n(\ve,p_z)=\frac{\ve+v\bm{\sigma}\cdot \mathbf{p}_n}{(\ve+i0)^2-\ve_n^2}, \\
           &x_{p_y}=x-p_y l_H^2.\ \
    \end{split}
\end{gather}
Here, $\chi_n\big(s\big)$ is the normalized oscillator wave function of the $n$th state and
\begin{gather}
\mathbf{p}_n=(0,\sqrt{2n}/l_H,p_z)
\end{gather}
is the effective 2D momentum.

\begin{figure}[h]
\centering
 \includegraphics[width=0.5\textwidth]{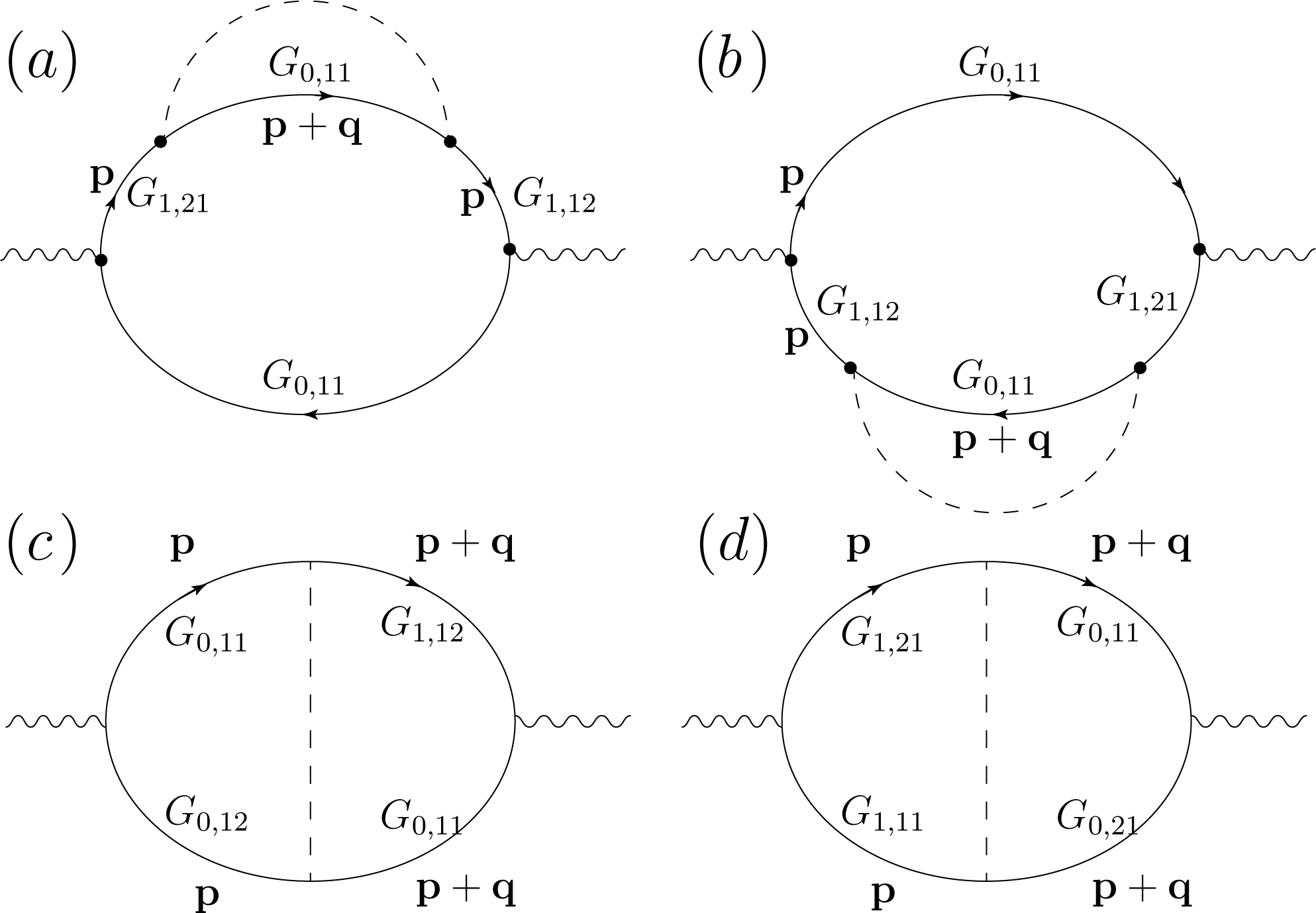}
 \caption{First-order contributions to the conductivity $\sigma_{xx}$. }
\label{diag1}
\end{figure}

\subsection{Summation of diagrams}
In the ultraquantum limit $(T\rightarrow 0)$, the Fermi function derivative can be substituted by the $\delta$ function, $\de f(\ve) = -\delta(\ve-\mu)$, and the integration over the energy can be explicitly performed. We will be interested in the small chemical potential limit, $\mu\ll\Omega$. As a result, we discard $\mu$ in the further computation of $\sigma_{xx}$ (but we will keep it for the computation of $\sigma_{xy}$ to obtain a nonvanishing result). As well as in Abrikosov's study~\cite{Abrikosov1998}, only the zeroth and the first Landau levels contribute to the conductivity.

We need to sum up the diagrams shown in Fig.~\ref{diag1}.
The details of the derivation are presented in Appendix~\ref{main_log}.

In the leading log approximation (the precision of Abrikosov's calculation~\cite{Abrikosov1998}), the conductivity $\sigma_{xx}$ has the form
\begin{gather}
\label{cond1}
    \sigma_{xx} = \frac{\alpha^3}{\Omega^2}v_\parallel^3 n_{\rm imp}\big[\cos^2\Theta+\xi^{-2}\sin^2\Theta\big]\ln\frac{1}{\alpha}.
\end{gather}
As we see, the anisotropy manifests itself in the $\Theta$ dependence of Eq.~\eqref{cond1}. At $\xi=1$, the $\Theta$ dependence drops out, and Eq.~\eqref{cond1} reproduces the famous Abrikosov's result for the isotropic WSM~\cite{Abrikosov1998}.
However, conductivity~\eqref{cond1} still does not exhibit the $\Phi$ dependence. This is due to the the insufficient precision of the log approximation. The result can be improved by a more accurate computation of the corresponding integrals.

After simple but a rather cumbersome analysis, we arrive at the following expression (see Appendix~\ref{main_log})
\begin{gather}
\label{cond_sublog}
\begin{split}
\sigma_{xx} =  \frac{\alpha^3}{2\pi\Omega^2}v_\parallel^3 n_{\rm imp}\big[\cos^2\Theta+\xi^{-2}\sin^2\Theta\big]\\
\times \Bigg[
\ln\frac{4\pi\xi^2}{\alpha e^C(\xi+\eta)^2}
-2\frac{\xi\cos^2\Phi+\eta\sin^2\Phi}{\xi+\eta}  \Bigg],
\end{split}
\end{gather}
where $\eta$ is defined in Eq.~\eqref{magnet_str} and $C$ is is the Euler--Mascheroni constant.

\begin{figure}[h]
\centering
 \includegraphics[width=0.5\textwidth]{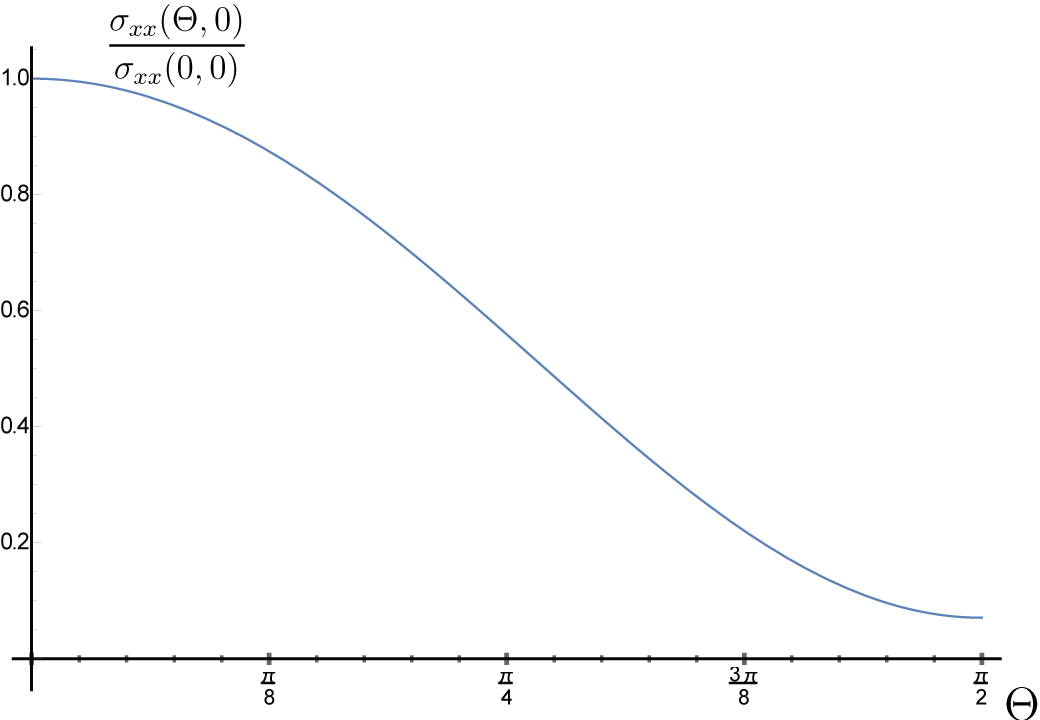}
 \caption{$\Theta$ dependence of the conductivity $\sigma_{xx}$ at $\Phi = 0$.}
\label{diag2}
\end{figure}

Expression~\eqref{cond_sublog} is the $\alpha$ expansion of the integrals entering Kubo formula~\eqref{cond2}, where the disorder averaging is performed with correlation function~\eqref{dis_res}. The omitted terms in the computation of integrals entering Kubo expression~\eqref{cond2} are of the order of $\mathcal{O}(\alpha\ln\alpha)$.
This is exactly the precision, with which we computed polarization operator in Coulomb potential~\eqref{coulomb2} and this makes the whole derivation to be selfconsistent.
The plots with the $\Theta$ and $\Phi$ dependence of magnetoconductivity are presented in Figs.~\ref{diag1} and \ref{diag2}.

\begin{figure}[h]
\centering
 \includegraphics[width=0.5\textwidth]{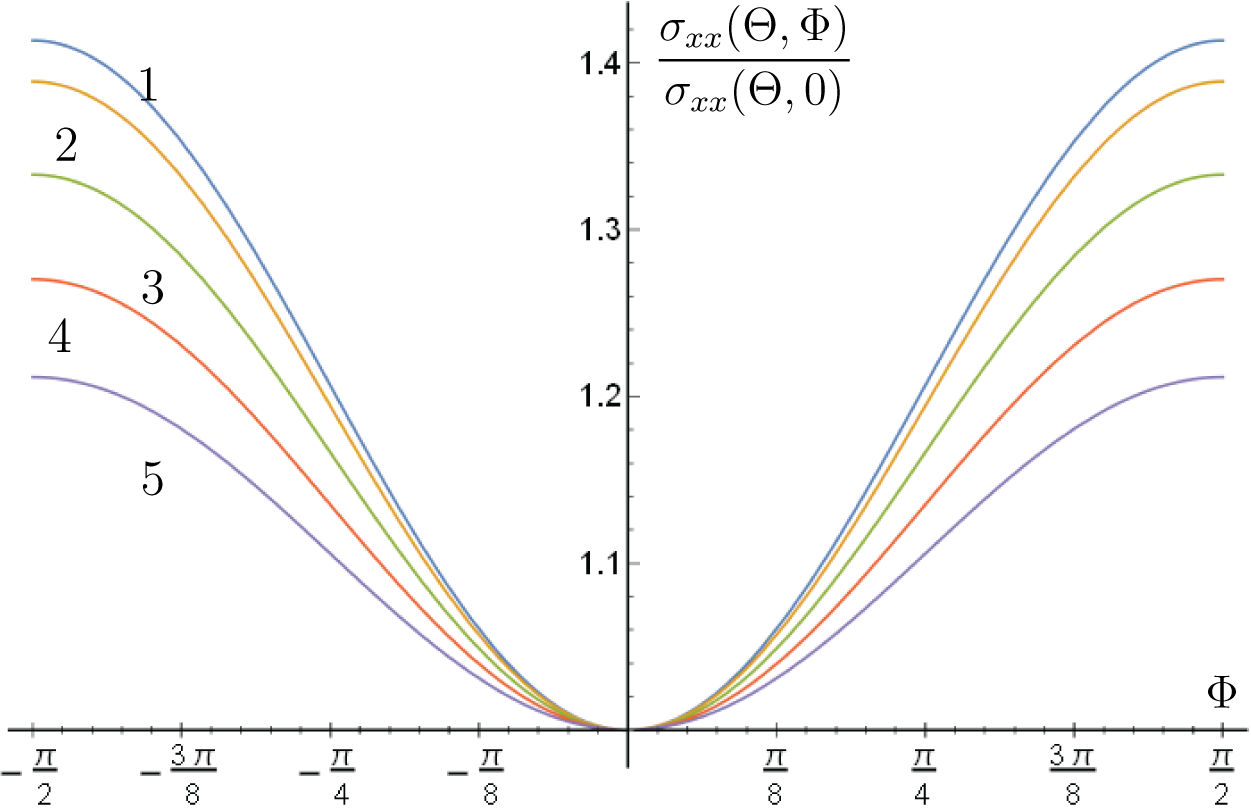}
 \caption{$\Phi$ dependence of the conductivity $\sigma_{xx}$ at different values of the polar angle $\Theta$. Here, $\Theta$ diminishes in $\pi/24$ steps (see curves from 1 to 5); $\Theta_n = \frac{\pi}{2}-(n-1)\frac{\pi}{24}$. The plots are drawn at the realistic values of $\xi = 4$ (Cd$_2$As$_3$) and of the fine structure constant $\alpha = 0.05$.}
\label{diag1}
\end{figure}

\section{Magnetoresistance}

The expression for magnetoresistance reads
\begin{gather}
\label{res1}
    \rho_{xx} = \frac{\sigma_{xx}}{\sigma_{xx}^2+\sigma_{xy}^2}.
\end{gather}
Two terms entering the denominator of Eq.~\eqref{res1} are not of the same order: $\sigma_{xx}$ is proportional to the disorder strength, while the first term in disorder expansion of $\sigma_{xy}$ is disorder independent. We are going to see that for not very highly compensated WSMs (see the exact condition below), the condition $\sigma_{xx}\ll\sigma_{xy}$ is always satisfied.

\subsection{Hall conductivity  $\sigma_{xy}$}

The Hall conductivity includes the anomalous and normal contributions. The full conductivity is disorder independent in the lowest order of the perturbation theory~\cite{Abrikosov1998,KlierPRB2017}. The expression, relating the Hall conductivity in the initial and rotated and rescaled basis follows from Eq.~\eqref{cond_fin_rot1} and reads
\begin{gather}
    \sigma_{xy}=\frac{1}{\xi}\sigma_{xy,sc}^\prime\sqrt{\xi^2\cos^2\Theta+\sin^2\Theta}.
\end{gather}
The expression for the Hall conductivity in the ultraquantum limit in the isotropic system can be taken from, e.g. Ref.~\onlinecite{Abrikosov1998}. We have
\begin{gather}
\label{hall_cond}
    \sigma_{xy} = \sqrt{\cos^2\Theta+\xi^{-2}\sin^2\Theta}\frac{\alpha\mu}{4\pi^2}.
\end{gather}

\subsection{Computation of the magnetoresistance}

We need to express the Hall conductivity~\eqref{hall_cond} via the charge carrier density. In the scaled rotated basis it is given by the standard expression~\cite{Abrikosov1998}: $n_{s} = \Omega_s^2\mu/(4\pi^2v_\parallel)$, where $\Omega_s^2 = 2eH_sv_\parallel/c$ is the magnetic field in the rescaled coordinate basis.

Using the relation between magnetic fields~\eqref{magnet_str}, we obtain the following relation for the charge carrier density
\begin{gather}
\label{density}
    n_0 = \frac{\Omega^2\mu}{4\pi^2 v_\parallel}\sqrt{\cos^2\Theta+\xi^{-2}\sin^2\Theta}.
\end{gather}
We see, that condition $\sigma_{xx}\ll\sigma_{xy}$ is met as long as $\alpha^2 n_{\rm imp}\ll n_0$. In the typical situation, the
electroneutrality condition entails $n_{\rm imp}\sim n_0$, therefore $\sigma_{xx}\ll\sigma_{xy}$ is always satisfied.

Finally, plugging in~\eqref{density} into~\eqref{hall_cond} and ~\eqref{res1}, we obtain the following expression for the magnetoresistance
\begin{gather}
\label{resist_fin}
\begin{split}
    \rho_{xx} = \frac{\Omega^2 n_{\rm imp} v_\parallel}{n_0^2}\big[\cos^2\Theta+\xi^{-2}\sin^2\Theta\big]\\
    \times
    \Bigg[
\ln\frac{4\pi\xi^2}{\alpha e^C(\xi+\eta)^2}
-2\frac{\xi\cos^2\Phi+\eta\sin^2\Phi}{\xi+\eta}  \Bigg].
\end{split}
\end{gather}
We see that the anisotropy is clearly pronounced in the realistic WSM where (like in Cd$_3$As$_2$, $\xi^2\approx 16\gg1$). If the anisotropy axis is oriented perpendicular to the magnetic field $\mathbf{H}$, the ratio of resistances scales as $\xi^2$
\begin{gather}
\label{resist}
    \frac{\rho_{xx}(\mathbf{H}\parallel\bn_0)}{\rho_{xx}(\mathbf{H}\bot\bn_0)} = \xi^2+\mathcal{O}\Big(\frac{1}{\ln\alpha}\Big).
\end{gather}
Expressions~\eqref{cond_sublog}, \eqref{resist_fin}, and \eqref{resist} are the main results of our paper.

It is quite interesting to point out once more that the azimuthal angle $\Phi$ dependence of the resistance manifests itself in the subleading to the main log term. Such is the consequence of averaging over long-range Coulomb disorder.

\section{Discussion}

We studied the magnetoresistance of WSM with an axial anisotropy. We found that the magnetoresistance is strongly renormalized as a function of the polar and azimuthal angle between anisotropy axis and the applied voltage plane. Some remarks are relevant here. First, we computed the contribution to conductivity from a single Weyl node. If the internodal scattering by disorder can be discarded, the total conductivity can be found by multiplying our Eq.~\eqref{resist_fin} by the number of Weyl nodes in the Brillouin zone. The internodal scattering can be neglected if the momentum transferred by disorder is much smaller then the distance between the adjacent Weyl nodes in momentum space. For Cd$_3$As$_2$  this distance reads\cite{krizman2019determination}: $2k_0 = 0.012$ \AA$^{-1}$, while for TaAs~\cite{lv2015observation} it is $2k_0 = 0.0183$ \AA$^{-1}$. On the other hand, the inverse Debye length in a typical magnetotransport experiment with field $H\sim 1$ T is $\kappa\sim 10^{-4}$ \AA$^{-1}$ (see Eq.~\eqref{coulomb2} and the comment below it). Therefore, indeed, we have $2k_0\gg\kappa$ and the internodal scattering can be safely neglected.

As was argued in Ref.~\onlinecite{Rodionov_PRB_15} at temperatures $T\gtrsim n_{\rm imp}^{1/3}v $, the electron--electron interaction starts dominating the transport in WSMs. For a typical magnetotransport experiment~\cite{jeon2014landau}, the charge carrier density $n\sim 10^{18}$ cm$^{-3}$, which yields the limiting temperature $T\lesssim 360$ K for the transport to be dominated by the Coulomb impurity scattering.
Therefore, we conclude that our findings should be valid in the majority of experiments dealing with magnetoresistance measurements in WSMs.

\section*{Acknowledgments}
The work was supported by the Russian Science Foundation (project No. 21-12-00254, https://rscf.ru/en/project/21-12-00254/). The work of A.S. Dotdaev in the part concerning the numerical calculations was supported by Grant No. K2-2022-025 in the framework of the Increase Competitiveness Program of NUST MISIS and by the the Foundation for the Advancement of Theoretical Physics and Mathematics ``Basis"  (project No. 22-1-1-24-1).

\appendix

\section{Derivation of the anisotropic Hamiltonian}
\label{App_ham}
In the rotated coordinate system, in which the $z^\prime$ axis is aligned along the anisotropy axis $\mathbf{n}_0$, we have the self-explanatory expression for the Hamiltonian
\begin{align}
    H_0 &= \int d\br^\prime \psi^{\prime\dag}(\br^\prime)
    h(p^\prime)\psi^{\prime}(\br^\prime)\notag\\
    h^\prime(p^\prime) &=
    \big[v_\bot(\sigma_x p_{x^\prime}+\sigma_y p_{y^\prime})+v_\parallel p_{z^\prime}\big],
    \label{ham_init}
\end{align}
where $\psi^\prime(\br^\prime)$ are spinors in the rotated basis.
We want to get back to the \textit{laboratory} system, in which the anisotropy axis $z^\prime$ ($\mathbf{n}_0$) is inclined at Euler angles $(\Phi,\Theta)$. The change from the laboratory system to the inclined is achieved by the rotation about the $z$ axis by angle $\Phi$  and about the new $y$ axis by $\Theta$. It is the standard Euler matrix relating vectors via $p = R p^\prime$, where
\begin{gather}
\label{rot}
    R= \left(
\begin{array}{ccc}
 \cos \Theta  \cos \Phi  & -\sin\Phi  & \sin \Theta  \cos \Phi  \\
 \cos \Theta  \sin \Phi  & \cos \Phi  & \sin \Theta  \sin \Phi  \\
 -\sin \Theta  & 0 & \cos \Theta  \\
\end{array}
\right).
\end{gather}
Therefore, the transformed Hamiltonian is rewritten as $h(p^\prime) \equiv h(R^{-1}p)$. We also need to transform the spinors according the 2D representation of the rotation group: $\psi^\prime = U\psi$, where the unitary matrix $U = U_{y}(\Theta)U_{z}(\Phi)$ is the product of unitary rotation matrices $U_\bn(\vf) = \exp[i\bm{\sigma}\cdot\bn\vf/2]$ by angle $\Phi$ about $z$ and by $\Theta$ about the new $y$ axis
\begin{gather}
\label{2d_rep}
    U = \left(
\begin{array}{cc}
 e^{\frac{i \Phi }{2}} \cos \frac{\Theta }{2} & e^{-\frac{i \Phi }{2} } \sin \frac{\Theta }{2} \\
 -e^{\frac{i \Phi }{2}} \sin \frac{\Theta }{2} & e^{-\frac{i\Phi}{2}} \cos \frac{\Theta }{2} \\
\end{array}
\right).
\end{gather}

Taking into account the fact that the Jacobian of the rotation is equal to unity ($d\br^\prime = d\br$), we see that the  transformed Hamiltonian~\eqref{ham_init} becomes
\begin{gather}
\begin{split}
    H_0 &= \int d\br \psi^\dag(\br)
    h(p)U\psi(\br),\\
    h(p) &=
    U^\dag h(R^{-1}p)U.
    \end{split}
\end{gather}
Taking expression~\eqref{ham_init} for the Hamiltonian $2\times2$ matrix $h^\prime(p^\prime)$ and performing direct substitution of $R$  from~\eqref{rot} and multiplication by matrices~\eqref{2d_rep}, we arrive at expression~\eqref{ham1}.

\section{Transformation of the conductivity tensor}
\label{App_cond_transform}

As was pointed out in Section~\ref{cond_transform}, the transformation due to rotation of the conductivity tensor is achieved via transform~\eqref{cond_rot1} with matrix~\eqref{rot}.
 The rescaling of coordinates $z = z^\prime,\ (x,y) = \xi(x^\prime,y^\prime)$ leads to the volume measure transform: $d\br = \xi^2 d\br^\prime$.
We require that the particle number be given by the same expression
\begin{gather}
\label{particle_num}
N = \int \psi^{\prime\dag}(\br^\prime)\psi^\prime(\br^\prime)d\br^\prime
\end{gather}
as before the rescaling. Hence, we postulate the scaling of the $\psi$ in such a way that expression~\eqref{particle_num} remains invariant
\begin{gather}
    \int d\br\psi^\dag(\br)\psi(\br) = \int\psi^\dag(Rr^\prime)\psi(Rr^\prime)\xi^2d\br^\prime,
\end{gather}
which entails the $\psi$-operator scaling law~\eqref{scaling}.

To understand how the conductivity tensor is transformed under the rescaling, we need to write the transformations for the electric field and current density. The electric field transformation law can be found from the requirement that the part of the Hamiltonian responsible for the coupling to the external electromagnetic field should remain unchanged (since the very definition of the conductivity is the response of the current to the external potential).

The corresponding potential affected by the rescaling enters the transversal part of canonical momentum and reads
\begin{gather}
\label{aux1}
-i\Big(\frac{\de}{\de \br_{\bot}}-\frac{ie}{c}\bA_\bot\Big) = -\frac{i}{\xi}\Big(\frac{\de}{\de \br_{s,\bot}}-i\frac{e}{c}\xi\bA_\bot\Big).
\end{gather}
Looking at~\eqref{aux1}, we immediately establish the scaling transformation for the vector potential
\begin{gather}
\label{vector_pot_sc}
    \bA_{s,\bot} = \xi \bA_\bot,\quad \bA_{s,\parallel}  =\bA_\parallel.
\end{gather}
Using definition $\mathbf{E} = -c^{-1}\de_t\bA,\  \mathbf{H}=\nabla\times\bA$,
we find transformation scaling rules for the electric and magnetic fields
\begin{gather}
\begin{split}
\label{electric_field}
    \mathbf{E}_{s,\bot} &= \xi \mathbf{E}_\bot,\quad
    \mathbf{E}_{s,\parallel} = \mathbf{E}_\parallel.\\
    \mathbf{H}_{s,\bot} &= \xi \mathbf{H}_\bot,\quad
    \mathbf{H}_{s,\parallel} = \xi^2\mathbf{H}_\parallel.
\end{split}
\end{gather}
Similarly, for the electric field, we find from~\eqref{electric_field}:
\begin{gather}
\label{matrix_S}
   E = S E_s,\ S\equiv  {\rm diag}(\xi^{-1},\xi^{-1},1).
\end{gather}
Finally, we determine the current density transformation law from its definition: $\mathbf{j} = n\mathbf{v}$. Recalling operator definition of the density and using~\eqref{scaling}, we write $n_s = \psi^{\dag}_s\psi_s = \xi^2n$. For the current density, we obtain
\begin{gather}
\label{current}
    \begin{split}
    \mathbf{j}_{s,\bot} &= \xi \mathbf{j}_\bot,\quad
    \mathbf{j}_{s,\parallel} = \xi^2\mathbf{j}_\parallel\ \Rightarrow\\
    j &= S_1 j_s,\quad S_1 = {\rm diag}(\xi^{-1}, \xi^{-1}, \xi^{-2}).
   \end{split}
\end{gather}
Now, from the definition of the conductivity tensor:
$j_i=\sigma_{ik}E_k$ with the help of Eqs.~\eqref{electric_field} and \eqref{current}, we determine the transformation law relating initial and scaled conductivities
\begin{gather}
    S_1 j_s = \sigma S E_s\ \Rightarrow \sigma_{s} = S_1^{-1}\sigma S\ \Rightarrow
    \sigma = S_1\sigma_{s} S^{-1}.
\end{gather}

\begin{widetext}
\section{Analytical expressions for diagrams, beyond the log approximation}
\label{main_log}

\subsection{Expression for the conductivity}
\label{exprcond}

In this section, the parameters $l_H$ and $\Omega^2$ refer to the rescaled quantities. Using the orthogonality relations for the Hermite polynomials, one easily convinces oneself that only the zeroth and the first LLs yield nonexponentially suppressed expressions corresponding to diagrams (a)--(d) represented in Fig.~\ref{diag1}
\begin{gather}
\label{diag_a}
    \begin{split}
    &(a):\quad \int\frac{dp_z }{2\pi}  {\rm Im}G_{0,11}(p_z)
    G^R_{1,21}(p_z) G^R_{1,12}(p_z)\int\frac{d\bq}{(2\pi)^3}{\rm Im}G_{0,11}^R(p_z+q_z) S_0(\bq)g(\bq), \\
    &S_0(\bq) = \int \frac{dp_y dx_1dx_2dx^\prime}{2\pi}e^{iq_x(x_1-x_2)} \chi^2_0(x_{p_y})\chi^2_0(x^\prime_{p_y})\chi_1(x_{1,p_y})\chi_0(x_{1,p_y+q_y})\chi_0(x_{2,p_y+q_y})
    \chi_1(x_{2,p_y}).
    \end{split}
\end{gather}
The diagrams in Figs.~\ref{diag1}(a) and \ref{diag1}(b) lead to identical expressions. The diagrams in Figs.~\ref{diag1}(c) and ~\ref{diag1}(d) read
\begin{gather}
\label{diag_c}
    \begin{split}
    &(c):\quad \int\frac{dp_z}{2\pi} \int\frac{d\bq}{(2\pi)^3}{\rm Im}G_{0,11}(p_z)
    {\rm Im}G_{0,11}^R(p_z+q_z) G_{1,12}(p_z+q_z) G_{1,12}(p_z) S_1(\bq) g(\bq),
    \\
    &S_1(\bq)=
    \!\!\int \frac{dp_y dx_1dx_2 dx^\prime}{2\pi} \chi^2_0(x_{p_y}) \chi^2_0(x^\prime_{p_y+q_y})  \chi_0(x_{1,p_y})\chi_1(x_{1,p_y+q_y}) e^{iq_x(x_1-x_2)}\chi_0(x_{2,p_y+q_y})
    \chi_1(x_{2,p_y}).
    \end{split}
\end{gather}
\begin{gather}
\label{diag_d}
    \begin{split}
    &(d):\quad \int\frac{dp_z}{2\pi} \int\frac{d\bq}{(2\pi)^3}{\rm Im}G_{0,11}(p_z)
    {\rm Im}G_{0,11}^R(p_z+q_z) G_{1,21}(p_z+q_z) G_{1,21}(p_z) S_2(\bq) g(\bq),
    \\
    &S_2(\bq)=
    \!\!\int \frac{dp_y dx_1dx_2 dx^\prime}{2\pi} \chi^2_0(x^\prime_{p_y}) \chi^2_0(x_{p_y+q_y})  \chi_0(x_{1,p_y})\chi_1(x_{1,p_y+q_y}) e^{-iq_x(x_1-x_2)}\chi_0(x_{2,p_y+q_y})
    \chi_1(x_{2,p_y}).
    \end{split}
\end{gather}
The expressions for form-factors $S_{0,1,2}(\bq)$ are easily computed using the relations for the Hermite polynomials. We have
\begin{gather}
    S_0(\bq) = \frac{1}{4\pi } e^{-q_\bot^2 l_H^2/2}q_\bot^2,\quad S_{1,2}(\bq) = \frac{1}{4\pi }e^{-q_\bot^2 l_H^2/2\pm2i\vf}q_\bot^2,
\end{gather}
where $\bq_\bot  = (q_x,q_y)$ and $\vf$ is its direction in the  $xy$ plane.

We also see that due to the presence of ${\rm Im} G_{0,11}(p_z) = -\pi\delta(v_\parallel p_z)$ and ${\rm Im} G_{0,11}(p_z+q_z)$, the integration over momenta $p_z$ and $q_z$ is trivial, leading effectively $p_z=q_z=0$.
As a result, the expressions for diagrams (a)--(d) are simplified
\begin{gather}
\label{abcd}
    (a+b+c+d)=\frac{1}{4\Omega^2}\frac{1}{4\pi
}\int\frac{d\bq_\bot}{(2\pi)^2}g(\bq_\bot)e^{-q_\bot^2 l_H^2/2}q_\bot^2\big(2-e^{2i\vf}-e^{-2i\vf}\big).
\end{gather}
Here, $g(\bq_\bot)$ is the potential correlation function taken at momentum $q_z = 0$. Using Eq.~\eqref{dis_res}, we write
\begin{gather}
    g(\bq_\bot) = \frac{16\pi^2n_{\rm imp}\xi^2\alpha^2v_\parallel^2}{\big(q_\bot^2[(\xi^2-1)\sin^2\gamma\cos^2(\vf-\vf_0)+1]+\xi^2\kappa^2\big)^2} .
\end{gather}
Here, $\vf_0$ is the azimuthal angle of the anisotropy axis. We see that this angle is in fact equal to zero in the rotated coordinate basis, since it belongs to the $x^\prime z^\prime$ plane. However, we will keep it arbitrary since we are going to need it for the computation of $\sigma_{yy}$ component.

Next, we are able to perform exactly the integration over $\vf$ and integration over $q_\bot$.

To go beyond the leading log approximation, we need to perform exact integration over $\vf$ in~\eqref{abcd}.
We use the following suitable integrals
\begin{gather}
    \int\limits_{-\pi}^\pi\frac{d\vf}{(\cos^2\vf+a^2)^2} = \frac{\pi}{a^3}\frac{2a^2+1}{(a^2+1)^{3/2}},\quad
        \int\limits_{-\pi}^\pi\frac{d\vf\cos2\vf}{(\cos^2\vf+a^2)^2} = -\frac{\pi}{a^3}\frac{1}{(a^2+1)^{3/2}},\quad a^2 = \frac{q_\bot^2+\xi^2\kappa^2}{q_\bot^2(\xi^2-1)\sin^2\gamma}.
\end{gather}
Then, we have for the~\eqref{abcd}
\begin{gather}
    (a+b+c+d) = \frac{n_{\rm imp}\xi^2\alpha^2}{\Omega^2(1+(\xi^2-1)\sin^2\gamma)^{3/2}}\int\limits_0^\infty q^3_\bot dq_\bot e^{-q_\bot^2 l_H^2/2}\frac{q_\bot^2\big[1+(\xi^2-1)\sin^2\gamma\cos^2\vf_0\big]
    +\xi^2\kappa^2}{(q_\bot^2+\kappa^2\xi^2)^{3/2}\big(q_\bot^2
    +\frac{\kappa^2\xi^2}{1+(\xi^2-1)\sin^2\gamma}\big)^{3/2}}.
\end{gather}

Let us introduce new integration variable: $q_\bot^2 = q_0^2 s,\ q_0^2 = \frac{\xi^2\kappa^2}{1+(\xi^2-1)\sin^2\gamma}$. Using relation $\kappa^2 l_H^2 = 2\alpha/(\pi\xi^2)$ (in the exponential function), and the handy relation $1+(\xi^2-1)\sin^2\gamma = \xi^2/\eta^2$, we arrive at the following dimensionless and convenient to analyze integral
\begin{gather}
\label{int1}
    (a+b+c+d) = \frac{n_{\rm imp}\eta^3\alpha^2}{2\xi \Omega^2}
    I(\vf_0),\quad I(\vf_0) = \int\limits_{0}^\infty\frac{s[1+(\xi^2-1)\sin^2\gamma\cos^2\vf_0]
    +\xi^2/\eta^2}{(s+1)^{3/2}(s+\xi^2/\eta^2)^{3/2}}s e^{-\alpha\eta^2 s/(\pi\xi^2)}\,ds.
\end{gather}
The integral in~\eqref{int1} can be computed for any value of $\vf_0$. However, we are going to need it at only two values: $\vf=0$ (for $\sigma^\prime_{s,xx}$) and for $\vf=\pi/2$ (for $\sigma^\prime_{s,yy}$). For brevity, let us denote $a = \xi/\eta\geq 1$. We are going to estimate them beyond log accuracy using the fact that $\alpha\ll 1$
\begin{gather}
\label{ints}
    I(0) = a^2\int\limits_{0}^\infty\frac{s}{(s+1)^{1/2}(s+a^2)^{3/2}}e^{-\alpha\eta^2 s/(\pi\xi^2)}\,ds,\quad
        I\Big(\frac{\pi}{2}\Big) = \int\limits_{0}^\infty\frac{s}{(s+1)^{3/2}(s+a^2)^{1/2}}e^{-\alpha\eta^2 s/(\pi\xi^2)}\,ds.
\end{gather}

For the conductivity, we have the following suitable expression
\begin{gather}
\label{aux3a}
    \sigma_{xx}= \frac{1}{a^2}\sigma^\prime_{s,xx}\cos^2\Phi+\sigma^\prime_{s,yy}\sin^2\Phi =\frac{\alpha^3 v_\parallel^3}{2\pi}\frac{n_{\rm imp}\eta^3}{\xi\Omega^2}\Big[\frac{1}{a^2}I(0)\cos^2\Phi+
    I\Big(\frac{\pi}{2}\Big)\sin^2\Phi)\Big].
\end{gather}

\subsection{Computation of the integrals}

In this case, both integrals entering~\eqref{ints} can be represented by the following expansion in $\alpha$: $\ln\frac{1}{\alpha}+{\rm const}+\mathcal{O}(\alpha)$. The integral accumulates its value on a span $s\lesssim \alpha^{-1}$. That means the momentum is $q\lesssim l_H^{-1}$. We are not interested in the $\mathcal{O}(\alpha)$ terms. However, we will extract const terms in both integrals since they carry the information on the $\Phi$ -- dependence of the conductivity. Both integrals can be represented as
\begin{gather}
\begin{split}
    &I(0) = a^2\big[J-a^2I_{0}(\alpha)\big],\quad I\Big(\frac{\pi}{2}\Big) = J-I_{\pi/2}(\alpha),\quad J = \int\limits_{0}^\infty\frac{1}{(s+1)^{1/2}(s+a^2)^{1/2}}e^{-\alpha\eta^2 s/(\pi\xi^2)}\,ds,\quad \\
    &I_0(\alpha) = \int\limits_{0}^\infty\frac{1}{(s+1)^{1/2}(s+a^2)^{3/2}}e^{-\alpha\eta^2 s/(\pi\xi^2)}\,ds,\quad
    I_{\pi/2}(\alpha)=\int\limits_{0}^\infty\frac{1}{(s+1)^{3/2}(s+a^2)^{1/2}}e^{-\alpha\eta^2 s/(\pi\xi^2)}\,ds.
\end{split}
\end{gather}
Both integrals $I_0(\alpha)$ and $I_{\pi/2}(\alpha)$ has regular limits at $\alpha\rightarrow 0$. Since we are not interested in $\mathcal{O}(\alpha)$ terms, we may set $\alpha = 0$ in them. We immediately obtain
\begin{gather}
\label{ints}
    I_0(0) = \frac{2}{a(a+1)},\quad I_{\pi/2}(0) = \frac{2}{(a+1)}.
\end{gather}
It is convenient to transform integral $J$ as
\begin{gather}
\label{aux2}
    J \equiv \int\limits_{0}^\infty\Big(\frac{1}{(s+1)^{1/2}(s+a^2)^{1/2}}-
    \frac{1}{s+1}\Big)e^{-\alpha\eta^2 s/(\pi\xi^2)}\,ds +\int\limits_{0}^\infty\frac{1}{s+1}e^{-\alpha\eta^2 s/(\pi\xi^2)}\,ds.
\end{gather}
The first term in~\eqref{aux2} is the convergent one, and one sets $\alpha = 0$. This gives $\ln(4(a+1)^{-2})$. The second integral is easily computed using the integration by parts. We obtain
\begin{gather}
\label{aux2}
    J  = \ln\frac{4\pi\xi^2}{(a+1)^2\alpha\eta^2 e^C}+\mathcal{O}(\alpha),
\end{gather}
where $C$ is the Euler--Mascheroni constant. 

Finally, let us deal with the the conductivity tensor. Using expression~\eqref{cond_fin_rot2} from the main body of the paper, and changing the rescaled $\Omega^2-\rightarrow \Omega^2\xi\eta$, we write
\begin{gather}
\label{aux3}
    \sigma_{xx}=\frac{\alpha^3 v_\parallel^3}{2\pi}\frac{n_{\rm imp}\eta^2}{\xi^2\Omega^2}\Big[(J-a^2I_0(0))\cos^2\Phi+(J-I_{\pi/2}(0)\sin^2\Phi)\Big].
\end{gather}

Plugging in~\eqref{ints} into~\eqref{aux3}, we obtain expression~\eqref{cond_sublog} for the conductivity.

\end{widetext}

\bibliographystyle{apsrevlong_no_issn_url}

\bibliography{aniso2}

\end{document}


\section{On the transformation of the conductivity tensor}.
First, let us recall that tensor turns vector into vector:
$v_\bot = \xi v_\parallel$.

\begin{gather}
    a = \sigma b
\end{gather}
Here, $a,\ b$ - vector components, $\sigma$ -components of the tensor.

Suppose, we change the coordinate basis. 
\begin{gather}
    b = R b^\prime,\quad a = R a^\prime,
\end{gather}
where $R$ is the transformation matrix. 
Then we have:
\begin{gather}
    R a^\prime = \sigma R b^\prime\ \rightarrow a^\prime = R^{-1}\sigma R b^\prime.
\end{gather}
Hence the tensor components in the transformed basis read:

\begin{gather}
    \sigma^\prime = R^{-1}\sigma R\ \Rightarrow\ \sigma  = R\sigma^\prime R^{-1}
\end{gather}

The transformation matrix from one basis to the other reads:
\begin{gather}
R = 
\left(
\begin{array}{ccc}
 \cos (\Theta ) \cos (\Phi ) & -\sin (\Phi ) & \sin (\Theta ) \cos (\Phi ) \\
 \cos (\Theta ) \sin (\Phi ) & \cos (\Phi ) & \sin (\Theta ) \sin (\Phi ) \\
 -\sin (\Theta ) & 0 & \cos (\Theta ) \\
\end{array}
\right)
\end{gather}
This matrix rotates the system in such a way that the new $z$-axis is aligned along the anisotropy vector with angular coordinates $(\Phi,\ \Theta)$.

Let us check: the unit vector along the new z-axis will have the components $n = (0,0,1)$. In the initial basis its components are given by the latter column of $R$ - matrix and indeed we see that they are correct.

\section{Rescaling of the conductivity}
Next we rescale coordinates according to $z = z^\prime,\ (x,y) = \xi(x^\prime,y^\prime)$.

To figure out how the conductivity transforms, let us recall that it relates current density and electric field vectors:

\begin{gather}
    j_i = \sigma_{ij}E_j.
\end{gather}
First, let us try to understand how the electric field is transformed. To this end we should understand how the scalar potential is affected by the scaling transformation. As we remember the $\psi$-operators are transformed according to:
\begin{gather}
    \psi(\br) = \frac{1}{\xi}\psi^\prime(\br^\prime)
\end{gather}

The scalar potential enters the Hamiltonian in the form:
\begin{gather}
    H = -\int\vf(\br)\psi^\dag(\br)\psi(\br)d\br = \int \vf\big[\br(\br^\prime)\big]\psi^{\prime\dag}(\br^\prime)\psi^\prime(\br^\prime)d\br^\prime
\end{gather}
Therefore, the potential stays invariant under transformation.
\begin{gather}
    \vf^\prime(\br^\prime) = \vf\big[\br(\br^\prime)\big].
\end{gather}

We have:
\begin{gather}
   \mathbf{E} =  
   \begin{pmatrix}
        E_x\\
        E_y\\
        E_z
    \end{pmatrix} = 
    \begin{pmatrix}
        \frac{\de\vf}{\de x}\\
        \frac{\de\vf}{\de y}\\
        \frac{\de\vf}{\de z}
    \end{pmatrix} =
    \begin{pmatrix}
        \frac{1}{\xi}\frac{\de\vf}{\de x^\prime}\\
        \frac{1}{\xi}\frac{\de\vf}{\de y^\prime}\\
        \frac{\de\vf}{\de z^\prime}
    \end{pmatrix}=
        \begin{pmatrix}
        \frac{1}{\xi}E_{x^\prime}\\
        \frac{1}{\xi}E_{y^\prime}\\
        E_{z^\prime}
    \end{pmatrix} = 
    \begin{pmatrix}
    \frac{1}{\xi}\  &    0\             &   0\\
         0\         &    \frac{1}{\xi}\ &   0\\
         0\         &    0\             &   1\\
    \end{pmatrix}
    \begin{pmatrix}
        E_{x^\prime}\\
        E_{y^\prime}\\
        E_{z^\prime}
    \end{pmatrix}\equiv S\mathbf{E}^\prime
\end{gather}    
Now we need to understand how the current density is transformed. Let us recall that $\mathbf{j} = \rho\mathbf{v}$.
The density is transformed according to:
\begin{gather}
    \rho = \frac{dN}{dV} = \frac{dN}{\xi^2 dV^\prime} = \frac{1}{\xi^2}\rho^\prime
\end{gather}
The velocity vector:
\begin{gather}
    \mathbf{v} = \begin{pmatrix}
        v_x\\
        v_y\\
        v_z
    \end{pmatrix} = 
    \begin{pmatrix}
        \frac{dx}{dt }\\
        \frac{dy}{dt }\\
        \frac{dz}{dt }
    \end{pmatrix} =
    \begin{pmatrix}
        \xi\frac{dx^\prime}{dt }\\
        \xi\frac{dy^\prime}{dt }\\
        \frac{dz^\prime}{dt }
    \end{pmatrix}=
        \begin{pmatrix}
        \xi v_{x^\prime}\\
        \xi v_{y^\prime}\\
        v_{z^\prime}
    \end{pmatrix}
\end{gather}    
Hence, the current:
\begin{gather}
    \mathbf{j} = 
    \begin{pmatrix}
        \frac{1}{\xi} j_{x^\prime}\\
        \frac{1}{\xi} j_{y^\prime}\\
        \frac{1}{\xi^2} j_{z^\prime}
    \end{pmatrix}= 
    \begin{pmatrix}
    \frac{1}{\xi}\  &    0\             &   0\\
         0\         &    \frac{1}{\xi}\ &   0\\
         0\         &    0\             &   \frac{1}{\xi^2}\\
    \end{pmatrix}
    \begin{pmatrix}
        j_{x^\prime}\\
        j_{y^\prime}\\
        j_{z^\prime}
    \end{pmatrix}\equiv S_1\mathbf{j}^\prime
\end{gather}
As a result we have:
\begin{gather}
    S_1\mathbf{j}^\prime = \sigma S\mathbf{E}^\prime\ \Rightarrow \sigma_{sc} = S_1^{-1}\sigma S\ \Rightarrow
    \sigma = S_1\sigma_{sc} S^{-1}.
\end{gather}
\section{The transformation of the magnetic field vector after the scaling}
The vector potential enters the Hamiltonian in the rotates basis as:
\begin{gather}
    H = \int\psi^\dag  \Big[v_\parallel\Big(-i\frac{\de}{\de z}  - A_z(\br)\Big)\sigma_z+
    v_\bot\Big(-i\frac{\de}{\de x}  - A_x(\br)\Big)\sigma_x+
    v_\bot\Big(-i\frac{\de}{\de y}  - A_y(\br)\Big)\sigma_y
     \Big]\psi dxdydz
\end{gather}
After the rescaling we obtain:
\begin{gather}
    H = v_\parallel\int\psi^{\prime\dag}  \Big[\Big(-i\frac{\de}{\de z^\prime}  - A_z(\br)\Big)\sigma_z+
    \Big(-i\frac{\de}{\de x^\prime}  - \xi A_x(\br)\Big)\sigma_x+
    \Big(-i\frac{\de}{\de y^\prime}  - \xi A_y(\br)\Big)\sigma_y
     \Big]\psi^\prime dx^\prime dy^\prime dz^\prime 
\end{gather}

Therefore, we see that the correct scaling of the vector potential reads:
\begin{gather}
    \begin{pmatrix}
        A^\prime_x\\
        A^\prime_y\\
        A^\prime_z
    \end{pmatrix} =
    \begin{pmatrix}
        \xi A_x\\
        \xi A_y\\
        A_z
    \end{pmatrix}
\end{gather}

After the scaling the magnetic field is transformed:
\begin{gather}
\label{mag_scale}
    \begin{split}
        H_x =\frac{\de A_z}{\de y} - \frac{\de A_y}{\de z}  = \frac{\de A_z^\prime}{\xi\de y^\prime}-\frac{\de A_y^\prime}{\xi\de z^\prime} =\frac{1}{\xi}H_x^\prime,\\
        H_y =\frac{\de A_x}{\de z} - \frac{\de A_z}{\de x} = 
        \frac{1}{\xi}\frac{\de A^\prime_x}{\de z^\prime} - \frac{1}{\xi}\frac{\de A^\prime_z}{\de x^\prime}=\frac{1}{\xi} H_y^\prime;\\
        H_z =\frac{\de A_y}{\de x} - \frac{\de A_x}{\de y} = 
        \frac{1}{\xi^2}\frac{\de A^\prime_y}{\de x^\prime} - \frac{1}{\xi^2}\frac{\de A^\prime_x}{\de y^\prime}=\frac{1}{\xi^2} H_z^\prime;
    \end{split}
\end{gather}

In the rotated coordinate system the vertical magnetic field vector has the following projections:
\begin{gather}
    \mathbf{H} = \begin{pmatrix}
        -H\sin\Theta\\
        0\\
        H\cos\Theta
    \end{pmatrix}
\end{gather}
After the rescaling, the components will change according to~\eqref{mag_scale}:
\begin{gather}
    \mathbf{H}_s = \begin{pmatrix}
        -H\xi\sin\Theta\\
        0\\
        H\xi^2\cos\Theta
    \end{pmatrix}
\end{gather}
These are the projections of the vector inclined by angle 
\begin{gather}
    \alpha = -\arctan\frac{1}{\xi}\tan\Theta.
\end{gather}
in the rotated $(x^\prime z^\prime)$ plane. To make it vertical (and this is what we need) we need to perform the backward rotation by $\alpha$ angle in $(x^\prime z^\prime)$ plane.

The corresponding transformation reads:
\begin{gather}
    R_{\alpha} = \left(
\begin{array}{ccc}
 \cos (\alpha ) & 0 & \sin (\alpha ) \\
 0 & 1 & 0 \\
 -\sin (\alpha ) & 0 & \cos (\alpha ) \\
\end{array}
\right)
\end{gather}
Now we are ready to link the conductivity $\sigma^\prime_{sc}$ in this final basis with the initial one:
\begin{gather}
\label{cond_fin_rot}
    \sigma_{sc} = R_\alpha \sigma^\prime_{sc} R_\alpha^{-1}\ \Rightarrow\ \sigma^\prime = S_1\sigma_{sc}S^{-1} =  S_1 R_\alpha \sigma^\prime_{sc} R_\alpha^{-1} S^{-1}\ \Rightarrow \notag\\
    \sigma = R\sigma^\prime R^{-1}= R S_1 R_\alpha \sigma^\prime_{sc} R_\alpha^{-1} S^{-1} R^{-1}. 
\end{gather}
The last expression is the final formula for the conductivity.
The perturbative conductivity tensor has the followign form:
\begin{gather}
    \sigma_{sc}^\prime = 
    \left(
\begin{array}{ccc}
 \sigma_{xx} & \sigma_{xy} & 0 \\
 -\sigma_{xy} & \sigma_{yy} & 0 \\
 0 & 0 & \sigma_{zz} \\
\end{array}
\right)
\end{gather}
We managed to prove (see below) that in the first order of perturbation theory $\sigma_{xz},\ \sigma_{yz} = 0$.
The formula~\eqref{cond_fin_rot} is the final expression we are going to work with. Let us present $\sigma_{xx}$:

\begin{gather}
\label{cond_fin_rot2}
    \sigma_{xx} = \frac{1}{\xi^2}\Big(\sigma^\prime_{xx,sc}\cos^2\Phi+\sigma^\prime_{yy,sc}\sin^2\Phi +(\xi^2-1)[\sigma^\prime_{xx,sc}\cos^2\Phi\cos^2\Theta+\sigma^\prime_{yy,sc}\sin^2\Phi]\Big)
\end{gather}
Let us check formula~\eqref{cond_fin_rot2} in some limiting cases. If anisotropy is absent $\xi= 1$ we obtain:
\begin{gather}
    \sigma_{xx} = \Big(\sigma^\prime_{xx,sc}\cos^2\Phi+\sigma^\prime_{yy,sc}\sin^2\Phi \Big)\quad (isotropic\ case)
\end{gather}
The last epxression is perfectly normal, since it relates the conductivity computed in the rotated by $\Phi$ around $z$ axis basis and the initial one. If the conductivity is isotropic in the rotated basis, then $\sigma^\prime_{xx,sc} = \sigma^\prime_{yy,sc}$ and we recover the normal relation:
$\sigma_{xx} = \sigma_{xx,sc}^\prime$.

\subsection{On the renormalization of the Debye length}
The screening of the Coulomb potential is given by the polarization operator.

\subsection{Kubo-Streda formula}
\subsection{Useful integrals}

\begin{gather}
\begin{split}
    &\int\limits_0^\infty\frac{s e^{-\beta s}\,ds}{(s+1)^{3/2}(s+a^2)^{1/2}} = 
    \big[K_0\Big(\frac{\beta}{2}(a^2-1)\Big)(1+\beta)+\beta K_1\Big(\frac{\beta}{2}(a^2-1)\Big)\big]e^{\frac{\beta}{2}(a^2+1)}-\frac{e^\beta}{\sqrt{a^2-1}}\Big[4\Phi_1\Big(\frac{1}{2},\frac12,\frac32,-\frac{1}{a^2-1},-\beta\Big)\\
    &+\frac{4}{3(a^2-1)}\Phi_1\Big(\frac12,\frac32,\frac52,\frac{1}{a^2-1},-\beta\Big)+\frac83 \Phi_1\Big(\frac12,\frac12,\frac52;\frac{1}{a^2-1},-\beta\Big)\Big]
\end{split}
\end{gather}

\begin{gather}
    \begin{split}
    &\int\limits_0^\infty\frac{s e^{-\beta s}\,ds}{(s+1)^{1/2}(s+a^2)^{3/2}} =
    \frac{4 e^{\beta } }{3 \left(a^2-1\right)^{3/2}}\Phi_1\Big(\frac12,\frac32,\frac52;\frac{1}{a^2-1},-\beta\Big)\\
    &+e^{\frac{1}{2} \left(a^2+1\right) \beta } \left(a^2 \beta  \left(K_0\left(\frac{\beta}{2} \left(a^2-1\right)  \right)-K_1\left(\frac{\beta}{2} \left(a^2-1\right) \right)\right)+K_0\left(\frac{\beta}{2} \left(a^2-1\right)\right)\right)
\end{split}
\end{gather}

\begin{gather}
    \int\limits_0^\infty \frac{e^{-t}}{\sqrt{t}\sqrt{t+a^2\beta}} = 
    e^{\frac{a^2\beta}{2}}K_0\Big(\frac{a^2\beta}{2}\Big).
\end{gather}

\begin{gather}
    \int\limits_0^\infty \frac{e^{-t}}{\sqrt{t}(t+a^2\beta)^{3/2}} = 
  .e^{\frac{a^2\beta}{2}}\bigg[K_1\Big(\frac{a^2\beta}{2}\Big)-K_0\Big(\frac{a^2\beta}{2}\Big)\bigg]
\end{gather}
\begin{gather}
    \int\limits_0^\infty \frac{e^{-t}}{\sqrt{t+\beta}(t+a^2\beta)^{3/2}} = e^{\frac{a^2 \beta }{2}} K_0\left(\frac{a^2 \beta }{2}\right)+\frac{\sqrt{\pi  \beta }}{a}-\frac{2}{a}+\mathcal{O}\Big(\beta^{3/2}\Big)
\end{gather}

\begin{gather}
    \int\limits_0^\infty \frac{t e^{-t}}{\sqrt{t+\beta}(t+a^2\beta)^{3/2}} = e^{\frac{a^2 \beta }{2}} \bigg[(1+a^2\beta)K_0\left(\frac{a^2 \beta }{2}\right)-a^2\beta K_1\left(\frac{a^2 \beta }{2}\right)\bigg]+\mathcal{O}\Big(\beta^{3/2}\Big)
\end{gather}
\begin{gather}
\int\limits_0^\infty \frac{t e^{-t}dt}{(t+\beta)^{3/2}(t+a^2\beta)^{1/2}}=
e^{\frac{a^2 \beta }{2}} \left((1-\beta ) K_0\left(\frac{a^2 \beta }{2}\right)+\beta  K_1\left(\frac{a^2 \beta }{2}\right)\right)+\frac{\sqrt{\pi } \sqrt{\beta }-4}{a}
\end{gather}
\begin{gather}
    \int\limits_0^\infty dt e^{-t}\sqrt{\frac{t+\beta}{t+a^2\beta}} = 
   \frac{a^2\beta}{2} e^{\frac{a^2\beta}{2}}\bigg[K_1\Big(\frac{a^2\beta}{2}\Big)-K_0\Big(\frac{a^2\beta}{2}\Big)\bigg]+
   \frac{\beta}{2} e^{\frac{a^2\beta}{2}}K_0\Big(\frac{a^2\beta}{2}\Big)
\end{gather}

\begin{gather}
    \int\limits_0^\infty \frac{e^{-t}\sqrt{t}}{(t+a^2\beta)^{1/2}} = 
  \frac{a^2\beta}{2}e^{\frac{a^2\beta}{2}}\bigg[K_1\Big(\frac{a^2\beta}{2}\Big)-K_0\Big(\frac{a^2\beta}{2}\Big)\bigg]
\end{gather}

\begin{gather}
    \int\limits_0^\infty \frac{e^{-t}\sqrt{t}}{(t+a^2\beta)^{3/2}} = 
  e^{\frac{a^2\beta}{2}}\bigg[(1+a^2\beta)K_0\Big(\frac{a^2\beta}{2}\Big)-a^2\beta K_1\Big(\frac{a^2\beta}{2}\Big)\bigg]
\end{gather}

\begin{gather}
    \int\limits_0^\infty \frac{e^{-t}}{(t+\beta)^{1/2}(t+a^2\beta)^{1/2}} = 
    -2\sqrt{\frac{\beta}{a^2\beta}}+
   e^{\frac{a^2\beta}{2}} K_0\Big(\frac{a^2\beta}{2}\Big)
  +e^{\frac{a^2\beta}{2}}\frac{\beta}{2}\bigg[(K_0\Big(\frac{a^2\beta}{2}\Big)+ K_1\Big(\frac{a^2\beta}{2}\Big)\bigg]+\mathcal{O}(\beta^{3/2})
\end{gather}

\begin{gather}
    \int\limits_0^\infty \frac{e^{-t}\sqrt{t}}{(t+a^2\beta)^{5/2}} = 
  \frac{e^{\frac{a^2\beta}{2}}}{3}\bigg[(1+2a^2\beta)K_1\Big(\frac{a^2\beta}{2}\Big)-(3+2a^2\beta) K_0\Big(\frac{a^2\beta}{2}\Big)\bigg]
\end{gather}

\subsubsection{Checked, very precise}
\begin{gather}
\int_0^{\infty } \frac{e^{-t}}{\sqrt{\beta +t} \left(a^2 \beta +t\right)^{3/2}} \, dt\\
=
-\frac{2 \sqrt{\beta }}{\left(a^2 \beta \right)^{3/2}}+\frac{\beta  \left(e^{\frac{a^2 \beta }{2}} \left(a^2 \beta  K_0\left(\frac{a^2 \beta }{2}\right)-\left(a^2 \beta -2\right) K_1\left(\frac{a^2 \beta }{2}\right)\right)\right)}{2 a^2 \beta }+e^{\frac{a^2 \beta }{2}} \left(K_1\left(\frac{a^2 \beta }{2}\right)-K_0\left(\frac{a^2 \beta }{2}\right)\right)+\mathcal{O}(\beta^{3/2})
\end{gather}

\begin{gather}
    \int\limits_0^\infty\frac{dt te^{-t}}{(t+\beta)^{3/2}(t+\mu)^{1/2}}=-4\sqrt{\frac{\beta}{\mu}}+
    e^{\mu/2}K_0+\frac{3\beta}{2}e^{\mu/2}\Big(K_0+K_1\Big)+\mathcal{O}(\beta^{3/2})
\end{gather}

\begin{gather}
     \int\limits_0^\infty\frac{dt te^{-t}}{(t+\beta)^{1/2}(t+a^2\beta)^{3/2}}= e^{\frac{a^2 \beta }{2}} \left(\left(a^2 \beta +1\right) K_0\left(\frac{a^2 \beta }{2}\right)-a^2 \beta  K_1\left(\frac{a^2 \beta }{2}\right)\right)-\frac{1}{2} \beta  e^{\frac{a^2 \beta }{2}} \left(K_1\left(\frac{a^2 \beta }{2}\right)-K_0\left(\frac{a^2 \beta }{2}\right)\right)+\mathcal{O}\Big(\beta^{3/2}\Big).
\end{gather}